\documentclass[submission]{eptcs}
\providecommand{\event}{SOS 2011}

\usepackage[usenames,dvipsnames]{color}

\usepackage{tikz}
\usetikzlibrary{matrix,arrows}

\usepackage{prooftree}

\usepackage{hyperref,,amsmath,amssymb,upgreek,breakurl}

\newcommand\waitchar[1]{\$\,#1} 
\newcommand\stackof[1]{\mathop{\texttt{stack}} #1} 
\newcommand\heapname{\pi} 
\newcommand\sendrule{\textsc{Send}} 
\newcommand\sync{\textsc{Sync}}  

\newcommand\evolve[1]{\mathop\square #1} 
\newcommand\proc{M} 
\newcommand\ppre[1]{#1\mkern2mu.\mkern2mu } 
\newcommand\apre[1]{\$\mkern2mu#1\mkern2mu.\mkern2mu } 
\newcommand\procsend[1]{\overline{#1}} 

\newcommand\astep[1]{\stackrel{#1}{\longrightarrow}} 
\newcommand\Astep[1]{\stackrel{#1}{\Longrightarrow}} 
\newcommand\macrostep{\Rightarrow}

\newcommand\parcomp{\mathbin{\|}} 
\newcommand\pwpiname{{PW}$\pi$} 

\newcommand\gramto{\mathrel{::=}} 

\usepackage{theorem}
\theorembodyfont{\rm}
\newtheorem{definition}{Definition}[section]

\newtheorem{theorem}[definition]{Theorem}
\newtheorem{example}[definition]{Example}

\numberwithin{equation}{section}
\numberwithin{figure}{section}

\newcommand\dom[1]{\mathrm{dom}(#1)}
\newcommand\sem[1]{[\mkern-2.5mu[ #1 ]\mkern-2.5mu]} 

\newcommand\blue[1]{\textcolor{blue}{#1}}
\newcommand\step{\mathrel{\blue\to}}
\newcommand\steps{\mathrel{\step\cdots\step}}
\newcommand\matches{\mathrel{\blue{\downarrow}}}
\newcommand\kleene[1]{{#1}{}^{*}\mkern-1.0mu}
\newcommand\infern[3]{\begin{prooftree} #1 \justifies #2 \using(#3) \end{prooftree}}
\newcommand\infer[2]{\begin{prooftree} #1 \justifies #2  \end{prooftree}}
\newcommand\mpair[2]{#1 \mathrel{\blue ;} #2}
\newcommand\mconf[1]{\blue{\langle} #1 \blue{\rangle}}
\newcommand\ekw[3]{\mconf{\mpair{#1}{\mpair{#2}{#3}}}}

\newcommand\ctnw[4]{\mconf{\mpair{#1}{\mpair{#2}{\mpair{#3}{#4}}}}}

\newcommand\knode[1]{\mathop{\texttt{cont}} #1}

\newcommand\nullpointer{{\normalfont\texttt{null}}}
\newcommand\mklist[1]{[#1]} 
\newcommand\empstring{\varepsilon} 
\newcommand\epsexp{\mathtt{\upvarepsilon}} 
\newcommand\emplist{\mklist{\,}} 
\newcommand\cons[2]{#1\mkern3mu{:\mkern-1.5mu:}\mkern3mu#2} 
\newcommand\conc[2]{#1@#2} 

\newcommand\pwpi[2]{\mconf{\mpair{#1}{#2}}}
\newcommand\addunique[3]{\psi(#1, #2, #3)}

\newcommand\stringconc[2]{#1 \; #2}

\newcommand\breakspace{\\[1ex]} 

\numberwithin{equation}{section}

\title{Regular Expression Matching and Operational Semantics}

\author{Asiri Rathnayake \qquad\qquad Hayo Thielecke
\institute{University of Birmingham \\ Birmingham B15 2TT, United Kingdom}
}

\def\titlerunning{Regular expression matching and operational semantics}
\def\authorrunning{Rathnayake and Thielecke}

\renewcommand\blue[1]{#1} 

\begin{document}

\maketitle

\begin{abstract}
Many programming languages and tools, ranging from grep to the Java String library, contain regular expression matchers. Rather than first translating a regular expression into a deterministic finite automaton, such implementations typically match the regular expression on the fly. Thus they can be seen as virtual machines interpreting the regular expression much as if it were a program with some non-deterministic constructs such as the Kleene star. We formalize this implementation technique for regular expression matching using operational semantics. Specifically, we derive a series of abstract machines, moving from the abstract definition of matching to increasingly realistic machines. First a continuation is added to the operational semantics to describe what remains to be matched after the current expression. Next, we represent  the expression as a data structure using pointers, which enables redundant searches to be eliminated via testing for pointer equality. From there, we arrive both at Thompson's lockstep construction and a machine that performs some operations in parallel, suitable for implementation on a large number of cores, such as a GPU. We  formalize the parallel machine using process algebra and report some preliminary experiments with an implementation on a graphics processor using CUDA.
\end{abstract}

\section{Introduction}

Regular expressions form a minimalistic language of pattern-matching constructs. Originally defined in Kleene's work on the foundations of computation, they have become ubiquitous in computing. Their practical significance was boosted by Thompson's efficient construction~\cite{thompson1968} of a regular expression matcher based on the ``lockstep'' simulation of a Non-deterministic Finite Automaton (NFA), and the wide use of regular expressions in Unix tools such as grep and awk.

The regular expression matchers used in such tools differ in detail
from the implementation of regular expressions used in compiler construction for lexical analysis.
In compiling, lexical analyzers are typically built by constructing a Deterministic Finite Automaton (DFA), using one of the standard results of automata theory. The DFA can process input very efficiently, but its construction incurs an additional overhead before any input can be matched. Moreover, the DFA construction only works if the matching language really is a regular language, so that it can be recognized by a DFA. Many matching languages add constructs that take the language beyond what a DFA can recognize, for instance back references. (By abuse of terminology, such extended languages are sometimes still referred to as ``regexes''.)

Recently, Cox~\cite{coxregexptwo} has given a rational reconstruction of Thompson's classic NFA matcher in terms of virtual machines. In essence, a regular expression is interpreted on the fly, much as a program in an interpreted programming language. The interpreter is a kind of virtual machine, with a small set of instructions suitable for running regular expressions. For instance, the Kleene star $e^{*}$ gives a form of non-deterministic loop. Cox emphasizes that the virtual machine approach in the style of Thompson is both flexible and efficient. Once a basic virtual machine for regular expressions is set up, other constructs such as back-references can be added with relative ease. Moreover, the machine is much more efficient than other implementation techniques based on a more naive backtracking interpreter~\cite{coxregexpone}, which exhibit exponential run-time in some cases. Surprisingly, these
inefficient matchers are widely used in Java and Perl~\cite{coxregexpone}.

In this paper, we formalize the view of regular expression matchers as machines by using tools from programming language theory, specifically operational semantics. We do so starting from the usual definition of regular expressions and their meaning, and then defining increasingly realistic machines.

We first define some preliminaries and recall what it means for a string to match a regular expression
in Section~\ref{bigstep}; from our perspective, matching is a simple form of big-step semantics, and we aim to refine it into a small-step semantics.
To do so in Section~\ref{ekwmachine}, we  introduce a distinction between a current expression and its continuation. We then refine this semantics by representing the regular expression as a syntax tree using pointers in memory (Section~\ref{ewpimachine}). Crucially, the pointer representation allows us to compare sub-expressions by pointer equality (rather than structurally). This pointer equality test is needed  for the efficient elimination of redundant match attempts, which underlies the general lockstep NFA simulation presented in Section~\ref{lock}. We recover Thompson's machine as a sequential implementation of the lockstep construction
(Section~\ref{lockstepmachine}). Since the lockstep construction involves simulating many non-deterministic machines in parallel, we then explore a parallel version using some simple process algebra in Section~\ref{gpulockstep}. The parallel process semantics is then related to a prototype implementation we have written in CUDA~\cite{cudaintro} to run on a Graphics Processor Unit (GPU)
in
Section~\ref{cudaimplementation}. Section~\ref{conclusions} concludes with some future directions. The overall plan of the paper can be visualised as follows:
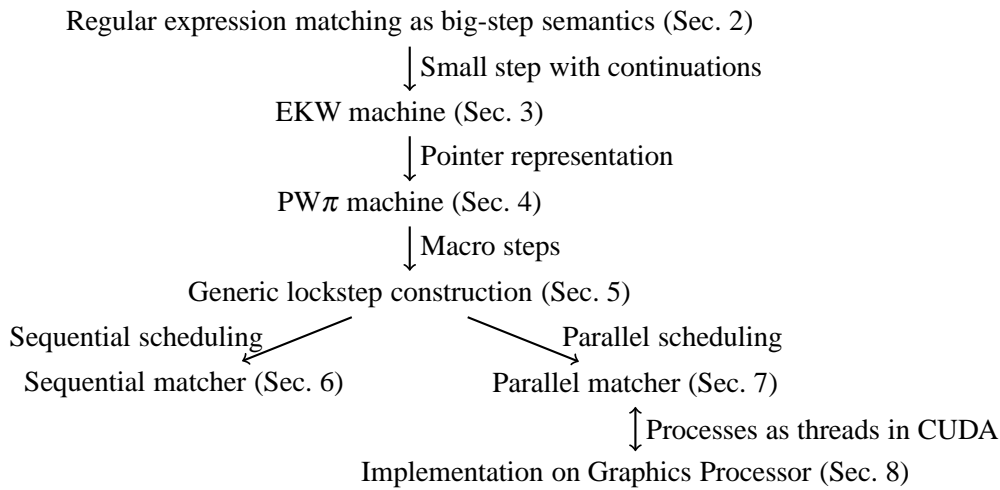
\[
\begin{tikzpicture}
\path (0, 0) 		node[rectangle] (regexp) {Regular expression matching as big-step semantics (Sec.~\ref{bigstep})}
 ++(0, -1.2) 		node[rectangle] (ekw) {EKW machine (Sec.~\ref{ekwmachine})}
 ++(0, -1.2) 		node[rectangle] (ewpi) {PW$\pi$ machine (Sec.~\ref{ewpimachine})}
 ++(0, -1.2) 		node[rectangle] (lockstep) {Generic lockstep construction (Sec.~\ref{lock})}
 ++(-3, -1.2) 		node[rectangle] (seq) {Sequential matcher (Sec.~\ref{lockstepmachine})}
 ++(+6, 0) 		node[rectangle] (par) {Parallel matcher (Sec.~\ref{gpulockstep})}
 ++(0, -1.2) 		node[rectangle] (cuda) {Implementation on Graphics Processor
 (Sec.~\ref{cudaimplementation})};

\path[->, thick] (regexp) edge node[auto] {Small step with continuations} (ekw) ;
\path[->, thick] (ekw) edge node[auto] {Pointer representation} (ewpi) ;
\path[->, thick] (ewpi) edge node[auto] {Macro steps} (lockstep) ;
\path[->, thick] (lockstep) edge node[left] {Sequential scheduling\ \ \ \ } (seq) ;
\path[->, thick] (lockstep) edge node[right] {\ \ \ \ Parallel scheduling} (par) ;
\path[<->, thick] (par) edge node[auto] {Processes as threads in CUDA} (cuda) ;
\end{tikzpicture}
\]


\section{Regular expression matching as a big-step semantics}
\label{bigstep}

Let $\Sigma$ be a finite set, regarded as the input alphabet.
We use the following abstract syntax for regular expressions:
\[
\begin{array}{rcll}
\renewcommand\breakspace{\\ }
e &\gramto& \epsexp \breakspace
e &\gramto& a  &\mbox{ where }a \in \Sigma\breakspace
e &\gramto& e^{*}\breakspace
e &\gramto& e_{1}e_{2} \breakspace
e &\gramto& e_{1} \mid e_{2}
\end{array}
\]

We let $e$ range over regular expressions, $a$ over characters, and $w$ over strings of characters. The empty string is written as $\empstring$. Note that there is also a regular expression constant $\epsexp$.
We also write the sequential composition $e_{1}\,e_{2}$ as $e_{1}\bullet e_{2}$ when we want to emphasise it as the occurrence of an operator applied to $e_{1}$ and $e_{2}$, for instance in a syntax tree.
For strings $w_{1}$ and $w_{2}$, we write their concatenation as juxtaposition $w_{1}w_{2}$. A single character $a$ is also regarded as a string of length $1$.

\begin{figure}
\fbox{$e \matches w$}
\[
\infern{e_{1} \matches w_{1} \qquad e_{2} \matches w_{2}}{(e_{1}\,e_{2}) \matches (w_{1}\,w_{2})}{\textsc{Seq}}
\qquad
\infern{}{a\matches a}{\textsc{Match}}
\qquad
\infern{}{\epsexp \matches \empstring}{\textsc{Epsilon}}
\]

\[
\infern{e\matches w_{1}\qquad \kleene e \matches w_{2}}{\kleene e \matches (w_{1}\,w_{2})}{\textsc{Kleene1}} \qquad \infern{}{\kleene e \matches \empstring}{\textsc{Kleene2}}
\]

\[
\infern{e_{1}\matches w}{(e_{1}\mid e_{2})\matches w}{\textsc{Alt1}} \qquad \infern{e_{2}\matches w}{(e_{1}\mid e_{2})\matches w}{\textsc{Alt2}}
\]
\caption{Regular expression matching as a big-step semantics}
\label{regexpmatch}
\end{figure}

Our starting point is the usual definition of what it means for a string $w$ to match a regular expression $e$. We write this relation as $e \matches w$, regarding it as a big-step operation semantics for a language with non-deterministic branching $e_{1} \mid e_{2}$ and a non-deterministic loop $\kleene e$. The rules are given in
Figure~\ref{regexpmatch}.

Some of our operational semantics will use lists.
We write $\cons{h}{t}$ for constructing a list with head $h$ and tail $t$. The concatenation of two lists $s$ and $t$ is written as \(\conc st\). For example, $\cons 1{[2]} = [1,2]$ and $\conc{[1,2]}{[3]} = [1,2,3]$. The empty list is written as $\emplist$.

\section{The EKW machine}
\label{ekwmachine}

The big-step operational semantics of matching in Figure~\ref{regexpmatch} gives us little information about how we should attempt to match a given input string $w$. We define a small-step semantics, called the EKW machine, that  makes the matching process more explicit.
 In the tradition of the SECD machine~\cite{landinmechanical}, the machine is named after its components: E for expression, K for continuation, W for word to be matched.

\begin{definition}
\label{defekw}
A configuration of the EKW machine is of the form $\ekw{e}{k}{w}$ where $e$ is a regular expression, $k$ is a list of regular expressions, and $w$ is a string. The transitions of the EKW machine are given in Figure~\ref{ekw}. The accepting configuration is $\ekw{\epsexp}{\emplist}{\empstring}$.

\begin{figure}
\renewcommand\breakspace{\\[.8ex]}
\fbox{$\ekw ekw \step \ekw{e'}{k'}{w'}$}
\begin{align}
\ekw{e_{1}\mid e_{2}}{k}{w}  &\step \ekw{e_{1}}{k}{w}\breakspace
\ekw{e_{1}\mid e_{2}}{k}{w}  &\step \ekw{e_{2}}{k}{w}\breakspace
\ekw{e_{1}\,e_{2}}{k}{w} &\step \ekw{e_{1}}{\cons{e_{2}}k}{w}\breakspace
\ekw{\kleene{e}}{k}{w}  &\step \ekw{e}{\cons{\kleene{e}}k}{w}\breakspace
\ekw{\kleene{e}}{k}{w}  &\step \ekw{\epsexp}{k}{w}\breakspace
\ekw{a}{k}{a\,w} &\step \ekw{\epsexp}{k}{w}\breakspace
\ekw{\epsexp}{\cons ek}{w} &\step \ekw{e}{k}{w}
\end{align}
\caption{EKW machine transition steps}
\label{ekw}
\end{figure}
\end{definition}
Here $e$ is the regular expression the machine is currently focusing on. What remains to the right of the current expression is represented by $k$, the current continuation. The combination of $e$ and $k$ together is attempting to match $w$, the current input string.

Note that many of the rules  are fairly standard, specifically the pushing and popping of the continuation stack. The machine is non-deterministic. The paired rules with the same current expressions $\kleene{e}$ or $(e_{1}\mid e_{2})$ give rise to branching in order to search for matches, where it is sufficient that one of the branches succeeds.

\begin{theorem}[Partial correctness]
\label{ekw_prop_partial}
$e\matches w$
if and only if there is a run
\[
\ekw{e}{\emplist}{w} \step \cdots \step \ekw{\epsexp}{\emplist}{\empstring}
\]
\end{theorem}

\begin{example}
\label{exampleekw}
Unfortunately, while Theorem~\ref{ekw_prop_partial} ensures that all matching strings are correctly accepted, there is no guarantee that the machine accepts all strings that it should on every run. In fact, there are valid inputs on which the machine may enter an infinite loop; an example is the configuration $\ekw{\kleene{\kleene a}}{\emplist}{a}$.
\begin{eqnarray*}
\ekw{\kleene{\kleene{a}}}{[\,]}{ a}
&\step &\ekw{\kleene{a}}{[\kleene{\kleene{a}}]}{ a}
\\
&\step &\ekw{\epsexp}{[\kleene{\kleene{a}}]}{ a}
\\
&\step &\ekw{\kleene{\kleene{a}}}{[\,]}{ a}
 \\
&\step &\cdots
\end{eqnarray*}
\end{example}
Such infinite loops can be prevented by backtracking and pruning. However, backtracking implementations can still take a very long time matching expressions like $\kleene{\kleene a}$ to a string consisting of, say, 1000 occurrences of a character $\texttt a$ followed by some other $\texttt b$, due to the exponentially increasing search space~\cite{coxregexpone}.

In Thompson's matcher, such loops are avoided by means of redundancy elimination. The matcher checks whether it has encountered the same expression before. Note, however, that ``the same'' expression is to be taken in the sense of pointer equality rather than structural equality. For instance, the two occurrences of $a$ in $(a\,b) \mid (a\,c)$ would be taken as not the same, given their different positions in the syntax tree.

\section{The \pwpiname{} machine}
\label{ewpimachine}

\begin{figure}[t]
\[
\begin{minipage}{\linewidth}\centering
\begin{tabular}{|c|c|c|}
\hline
$p$ & $\pi(p)$ & $\knode{p}$\\
\hline\hline
$p_{0}$ & $p_{1} \bullet p_{2}$ & $\nullpointer$\\
\hline
$p_{1}$ & $\kleene{p_{3}}$ & $p_{2}$\\
\hline
$p_{2}$ & $b$ & $\nullpointer$\\
\hline
$p_{3}$ & $\kleene{p_{4}}$ & $p_{1}$\\
\hline
$p_{4}$ & $a$ & $p_{3}$\\
\hline
\end{tabular}
\begin{minipage}[!t]{0.5\linewidth}\centering
\begin{tikzpicture}
\path (1, 1) 		node[rectangle] (null) {$\texttt{null}$};
\path (-0.4, 0.4)	node[rectangle] (b0) {$p_{0}$};
\path (0, 0) 		node[draw, circle] (p0) {$\bullet$};
\path (-1.4, -0.6)	node[rectangle] (b1) {$p_{1}$};
\path (-1, -1) 		node[draw, circle] (p1) {$*$};
\path (1.4, -0.6)	node[rectangle] (b2) {$p_{2}$};
\path (1, -1) 		node[draw, circle] (p2) {$b$};
\path (-1.4, -1.6)	node[rectangle] (b3) {$p_{3}$};
\path (-1, -2) 		node[draw, circle] (p3) {$*$};
\path (-1.4, -2.6)	node[rectangle] (b4) {$p_{4}$};
\path (-1, -3) 		node[draw, circle] (p4) {$a$};

\draw[->, thick](p0) -- (p1);
\draw[->](p0) edge[dashed] (null);
\draw[->, thick](p0) -- (p2);
\draw[->](p2) edge[dashed] (null);
\draw[->, thick](p0) -- (p2);
\draw[->](p1) edge[dashed] (p2);
\draw[->, thick](p1) -- (p3);
\draw[->](p3) edge[bend right=45, dashed] (p1);
\draw[->, thick](p3) -- (p4);
\draw[->](p4) edge[bend right=45, dashed] (p3);
\end{tikzpicture}\end{minipage}
\end{minipage}
\]
\caption{The regular expression $\kleene{\kleene a}\bullet b$ as a tree with continuation pointers}
\label{figurepi}
\end{figure}
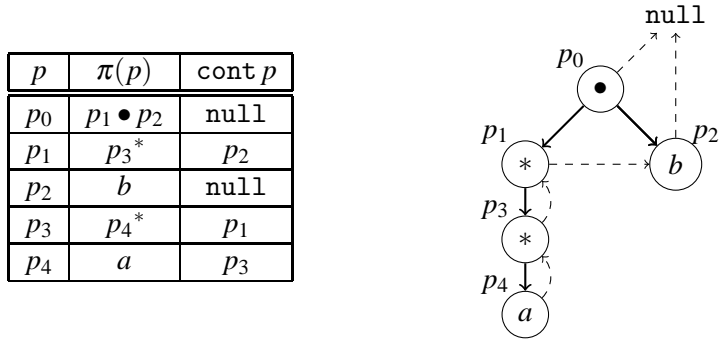

We refine the EKW machine by representing the regular expression as a data structure in a heap $\heapname$, which serves as the program run by the machine. That way, the machine can distinguish between different positions in the syntax tree.

\begin{definition}
A heap $\pi$ is a finite partial function from addresses to values. There exists a distinguished address $\nullpointer$, which is not mapped to any value.
\end{definition}
In our setting, the values are syntax tree nodes, represented by an operator from the syntax of regular expressions together with pointers to the tree for the arguments (if any) of the operator. For example,
for sequential composition, we have a node containing $(p_{1} \bullet p_{2})$, where the two pointers $p_{1}$ and $p_{2}$ point to the trees of the two expressions being composed.
\begin{definition}
We write $\otimes$ for the partial operation of forming the union of two partial functions provided that their domains are disjoint. More formally, let $f_{1}: A \rightharpoonup B$ and $f_{2}: A \rightharpoonup B$ be two partial functions. Then if $\dom{f_{1}} \cap \dom{f_{2}} = \emptyset$, the function
\[
(f_{1} \otimes f_{2}): A \rightharpoonup B
\]
is defined as $f_{1} \otimes f_{2} = f_{1} \cup f_{2}$.
\end{definition}

 Note that $\otimes$ is the same as the operation $*$ on heaps in separation logic~\cite{reynoldslicssep}, and hence a partial commutative monoid. We avoid the notation $*$ as it could be confused with the Kleene star. As in separation logic, we use $\otimes$ to describe data structures with pointers in memory.
\begin{definition}
We write $\pi,p \models e$ if $p$ points to the root node of a regular expression $e$ in a heap $\pi$. The relation is defined by induction on $e$ as follows:
\[
\begin{array}{llrl}
\renewcommand\breakspace{\\[.8ex]}
\pi, p &\models  a &&\mbox{if } \pi(p) = a\breakspace
\pi, p &\models  \epsexp &&\mbox{if } \pi(p) = \epsexp\breakspace
\pi, p &\models (e_{1} \mid e_{2}) &&\mbox{if } \pi = \pi_{0} \otimes \pi_{1} \otimes \pi_{2} 
\wedge \pi_{0}(p) = (p_{1}\mid p_{2})
\breakspace
&&&
\wedge \pi_{1},p_{1} \models e_{1}
\wedge \pi_{2},p_{2}\models e_{2}\breakspace
\pi, p &\models (e_{1} \, e_{2}) &&\mbox{if } \pi = \pi_{0} \otimes \pi_{1} \otimes \pi_{2}
\wedge \pi_{0}(p) = (p_{1}\bullet p_{2})\breakspace
&&&\wedge \pi_{1},p_{1} \models e_{1}
\wedge \pi_{2},p_{2}\models e_{2}\breakspace
\pi, p &\models \kleene{e_{1}} &&\mbox{if } \pi = \pi_{0} \otimes \pi_{1}
\wedge \pi_{0}(p) = \kleene{p_{1}}
\wedge \pi_{1},p_{1} \models e_{1}
\end{array}
\]
\end{definition}
\noindent
Here the definition of $\pi,p \models e$  precludes any cycles in the child pointer chain.

As an example, consider the regular expression $e = \kleene{\kleene{a}} b$. A $\pi$ and $p_{0}$ such that $\pi,p_{0} \models e$ is given by the table in
Figure~\ref{figurepi}. The tree structure, represented by the solid arrows, is drawn on the right.

\begin{definition}
\label{defknode}
Let $\knode$ be a function
\[
\knode{} : \dom{\pi} \to (\dom{\pi} \cup \{ \nullpointer \})
\]
We write $\pi \models \knode{}$ if
\begin{itemize}
\item
If $\pi(p) = (p_{1} \mid p_{2})$, then $\knode{p_{1}} = \knode{p}$ and $\knode{p_{2}} = \knode{p}$
\item
If $\pi(p) = (p_{1} \bullet p_{2})$, then $\knode{p_{1}} = p_{2}$ and $\knode{p_{2}} = \knode{p}$
\item
If $\pi(p) = \kleene{(p_{1})}$, then $\knode{p_{1}} = p$
\item
$\knode{p_{0}} = \nullpointer$, where $p_{0}$ is the pointer to the root of the syntax tree.
\end{itemize}
\end{definition}

The function $\knode$ is uniquely determined by the tree structure layed out in $\pi$, and it is easy to compute by a recursive tree walk. We elide it when it is clear from the context, assuming that $\pi$ always comes equipped with a $\knode$ such that $\pi\models\knode$.
By treating $\knode$ as a function, we have not committed to a particular implementation;
for instance
$\knode$ could be represented as a hash table indexed by pointer values, or it could be added as another pointer field to the nodes in the heap.

 In the graphical representation
in Figure~\ref{figurepi}, dashed arrows represent $\knode$. In particular, note the cycle leading downward from $p_{1}$ and up again via dashed arrows. Following such a cycle could lead to  infinite loops as for the EKW machine in
Example~\ref{exampleekw}.

\begin{figure}
\fbox{$p \astep{} q$ or $p \astep{a} q$ relative to $\pi$}
\[
\begin{array}{rcll}
p &\astep{} & p_{1}
&\textrm{ if } \pi(p) = p_{1} \mid p_{2}
\\[1ex]
p &\astep{} &  p_{2}
&\textrm{ if } \pi(p) = p_{1} \mid p_{2}
\\[1ex]
p &\astep{} &  p_{1}
&\textrm{ if } \pi(p) = p_{1} \bullet p_{2}
\\[1ex]
p &\astep{} &  p_{1}
&\textrm{ if } \pi(p) = \kleene{p_{1}}
\\[1ex]
p &\astep{} &  p_{2}
&\textrm{ if } \pi(p) = \kleene{p_{1}} \mbox{ and }\knode p = p_{2}
\\[1ex]
p &\astep{} &  p_{1}
&\textrm{ if } \pi(p) = \epsexp \mbox{ and }\knode p = p_{1}
\\[1em]
p &\astep a& p'
& \textrm{ if }
 \pi(p) = a \mbox{ and }p' = \knode p
\end{array}
\]
\caption{PW$\pi$ transitions}
\label{pwpistep}
\end{figure}

\begin{definition}
\label{pwpidef}
The \pwpiname{} machine is defined as follows.
Transitions of this machine are always relative to some heap $\pi$, which does not change during evaluation. We elide $\pi$ if it is clear from the context.
Configurations of the machine are of the form $\pwpi pw$, where $p$ is a pointer in $\pi$ and $w$ is a string of input symbols.
Given the transition relation between pointers defined in Figure~\ref{pwpistep}, the machine has the following transitions:
\[
\infer{p \astep{a} q}{\pwpi{p}{\stringconc aw} \step \pwpi{q}{w} }
\hspace{3em}
\infer{p \astep{} q}{\pwpi{p}{\stringconc w} \step \pwpi{q}{w} }{}
\]
The
accepting state of the machine is $\pwpi{\nullpointer}{\empstring}$.  That is, both the continuation and the remaining input have been consumed.
\end{definition}

\begin{example}
For a regular expression $e = \kleene{\kleene{a}} b$, let $\pi$ and $p_{0}$ be such that $\pi,p_{0} \models e$. See Figure~\ref{figurepi} for the representation of $\pi$ as a tree with pointers. The diagram below illustrates two possible executions of the PW$\pi$ machine against inputs $e$ and $aab$.
\\

\begin{minipage}[!t]{0.5\linewidth}\centering
Execution - 1: Infinite loop
\begin{eqnarray*}
&\pwpi{p_{0}}{aab}\\
\longrightarrow &\pwpi{p_{1}}{aab}\\
\longrightarrow &\pwpi{p_{3}}{aab}\\
\longrightarrow &\pwpi{p_{1}}{aab}\\
\longrightarrow &\pwpi{p_{3}}{aab}\\
\longrightarrow &\pwpi{p_{1}}{aab}\\
\longrightarrow &\pwpi{p_{3}}{aab}\\
\longrightarrow &\pwpi{p_{1}}{aab}\\
\longrightarrow &\pwpi{p_{3}}{aab}\\
\longrightarrow &\pwpi{p_{1}}{aab}\\
\longrightarrow &\ldots
\end{eqnarray*}
\end{minipage}
\begin{minipage}[!t]{0.5\linewidth}\centering
Execution - 2: Successful match
\begin{eqnarray*}
&\pwpi{p_{0}}{aab}\\
\longrightarrow &\pwpi{p_{1}}{aab}\\
\longrightarrow &\pwpi{p_{3}}{aab}\\
\longrightarrow &\pwpi{p_{4}}{aab}\\
\longrightarrow &\pwpi{p_{3}}{ab}\\
\longrightarrow &\pwpi{p_{4}}{ab}\\
\longrightarrow &\pwpi{p_{3}}{b}\\
\longrightarrow &\pwpi{p_{1}}{b}\\
\longrightarrow &\pwpi{p_{2}}{b}\\
\longrightarrow &\pwpi{\nullpointer}{\empstring}\\
\end{eqnarray*}
\end{minipage}
\label{pwpiexample}

\end{example}

\begin{theorem}[Simulation]
\label{simekwpwpi}
Let $\pi$ be a heap such that $\pi,p \models e$. Then there is a run of the EKW machine of the form
\[
\ekw{e}{\emplist}{w} \steps \ekw{\epsexp}{\emplist}{\empstring}
\]
if and only if there is a run of the $PW\pi$ machine of the form
\[
\pwpi{p}{w} \steps \pwpi{\nullpointer}{\empstring}
\]
\end{theorem}
One needs to show that each step of the EKW machine can be simulated by the \pwpiname{} machine and vice versa. The invariant in this simulation
 is that the stack $k$ in the EKW machine can be reconstructed by following the chain of pointers in the heap of the \pwpiname{} machine via the following function:
 \[
\begin{array}{rcll}
\stackof p
&=&
\emplist
&\mbox{if } \knode p = \nullpointer
\\
\stackof p
&=& e \cons {} (\stackof q)
&\mbox{if } q= \knode p  \neq \nullpointer
\\
&&&\mbox{and } \pi, q \models e
\end{array}
\]

\section{The lockstep construction in general}
\label{lock}
As we have seen, the \pwpiname{} machine is built from two kinds of steps. Pointers can be evolved via $p \astep{} q$ by moving in the syntax tree without reading any input. When a node for a constant is reached, it can be matched to the first character in the input via a step $p \astep a q$.
\begin{definition}
\label{evolved}
Let $S \subseteq\dom\pi\cup\{\nullpointer\}$ be a set of pointers. We define the evolution $\evolve S$ of $S$ as the following set:
\[
\evolve S
=
\{q \in \dom{\pi} \mid \exists p \in S. p\astep{}^{*} q \wedge \exists a.\pi(q) = a \}
\]
\end{definition}

Forming $\evolve S$ is similar to computing the $\varepsilon$-closure in automata theory.
However, this operation is not a closure operator, because $S \subseteq \evolve S$ does not hold in general.
 When one computes $\evolve S$ incrementally, elements are removed as well as added. Avoiding infinite loops by adding and removing the same element is the main difficulty in the computation.


We define a transition relation analogous to Definition~\ref{pwpidef}, but as a deterministic relation
on \emph{sets} of pointers. We refer to these as macro steps, as they assume the computation of $\evolve S$ as given in a single step, whereas an implementation needs to compute it incrementally.

\begin{definition}[Lockstep transitions]
\label{locktransdef}
Let $S, S'\subseteq\dom\pi\cup\{\nullpointer\}$ be sets of pointers.
\[
\begin{array}{rcll}
S & \Astep{} & S'
& \textrm{ if } S' = \evolve S \\[1em]
S & \Astep{a} & S'
& \textrm{ if } S' =  \{ q \in \dom{\pi} \mid \exists p \in S. p\astep{a} q \}
\end{array}
\]
\end{definition}
A set of pointers is first evolved from $S$ to $\evolve S$.
Then, moving from a set of pointers $\evolve S$ to $S'$ via $\evolve S \Astep{a} S'$  advances the state of the machine by advancing all pointers that can match $a$ to their continuations. All other pointers are deleted as unsuccessful matches.

\begin{definition}[Generic lockstep machine]
\label{genlockstepmachine}
The generic lockstep machine has configurations of the form $\pwpi Sw$. Transitions are defined using  Definition~\ref{locktransdef}:
\[
\infer{S \Astep{a} S'}{\pwpi{S}{\stringconc aw} \macrostep \pwpi{S'}{w} }
\hspace{3em}
\infer{S \Astep{} S'}{\pwpi{S}{\stringconc w} \macrostep \pwpi{S'}{w} }{}
\]
Accepting states of the machine are of the form $\pwpi  S\empstring$, where $\nullpointer\in S$.
\end{definition}

\begin{theorem}
\label{pwpithm}
For a heap $\pi, p \models e$ there is a run of the PW$\pi$ machine:
\[
\pwpi{p}{w} \steps \pwpi{\nullpointer}{\empstring}
\]
if and only if there is a run of the lockstep machine
\[
\pwpi{\{p\}}w \macrostep\ldots \macrostep \pwpi{S}{\empstring}
\]
for some set of pointers $S$
with $\nullpointer \in S$.
\end{theorem}

\section{The sequential lockstep machine}\label{lockstepmachine}
The sequential lockstep machine maintains two lists of pointers $c$, $n$ corresponding to pointers being incrementally evolved within the current macro step and pointers to be evolved in the next macro step. Another pointer list $t$ is maintained which provides support for redundancy elimination, we also introduce an auxilary function $\addunique{p}{l_{1}}{l_{2}}$ to aid in this regard:
\begin{definition}
The auxilary function $\addunique{p}{l_{1}}{l_{2}}$ is defined as:
\begin{align*}
&\addunique{p}{l_{1}}{l_{2}} = \cons{p}{l_{1}} \textrm{ if } p \notin \conc{l_{1}}{l_{2}}\\
&\addunique{p}{l_{1}}{l_{2}} = l_{1} \textrm{ if } p \in \conc{l_{1}}{l_{2}}
\end{align*}
\end{definition}
\begin{definition}
The redundancy-eliminating sequential lockstep machine has configurations of the form $\ctnw{c}{t}{n}{w}$. Its transitions are given in figure \ref{red_free_lock_step_trans}.
The accepting states are of the form $\ctnw{\cons{\nullpointer}{c'}}{t'}{n'}{\empstring}$ \begin{figure}
\fbox{$\ctnw{c}{t}{n}{w}\step\ctnw{c'}{t'}{n'}{w'}$}
\[
\begin{array}{rcll}
\renewcommand\breakspace{\\[.8em]}
\ctnw{\cons{p}{c}}{t}{n}{w} &\step& \ctnw{c'}{\cons{p}{t}}{n}{w} &\textrm{ if } \pi(p) = p' \mid p'' \notag
\\
&&&\textrm{ where } c' = \addunique{p''}{\addunique{p'}{c}{t}}{t}
\breakspace
\ctnw{\cons{p}{c}}{t}{n}{w} &\step& \ctnw{c'}{\cons{p}{t}}{n}{w} \notag &\textrm{ if } \pi(p) = p' \bullet p''
\\
&&&\textrm{ where } c' = \addunique{p'}{c}{t}
\breakspace
\ctnw{\cons{p}{c}}{t}{n}{w} &\step& \ctnw{c'}{\cons{p}{t}}{n}{w} \notag &\textrm{ if } \pi(p) = \kleene{(p')} \notag\\
&&&\textrm{ where } c' = \addunique{\knode{p}}{\addunique{p}{c}{t}}{t}
\breakspace
\ctnw{\cons{p}{c}}{t}{n}{w} &\step& \ctnw{c'}{\cons{p}{t}}{n}{w} \notag &\textrm{ if } \pi(p) = \epsexp
\\
&&&\textrm{ where } c' = \addunique{\knode{p}}{c}{t}
\breakspace
\ctnw{\cons{p}{c}}{t}{n}{a\,w} &\step& \ctnw{c}{t}{n}{a\,w} \notag &\textrm{ if } p = \nullpointer
\breakspace
\ctnw{\cons{p}{c}}{t}{n}{a\,w} &\step& \ctnw{c}{\cons{p}{t}}{n'}{a\,w} \notag &\textrm{ if } \pi(p) = a \notag\\
&&&\textrm{ where } n' =  \addunique{\knode{p}}{n}{\emplist}
\breakspace
\ctnw{\cons{p}{c}}{t}{n}{a\,w} &\step& \ctnw{c}{\cons{p}{t}}{n}{a\,w} \notag &\textrm{ if } \pi(p) = b
\breakspace
\ctnw{\emplist}{t}{n}{a\,w} &\step& \ctnw{n}{\emplist}{\emplist}{w} \notag &\textrm{ if } n \ne \emplist
\breakspace
\ctnw{\cons{p}{c}}{t}{n}{\empstring} &\step& \ctnw{c}{\cons{p}{t}}{n}{\empstring} \notag &\textrm{ if } \pi(p) = a
\end{array}
\]
\caption{Sequential lockstep machine with redundancy elimination}
\label{red_free_lock_step_trans}
\end{figure}
\end{definition}

We regard this machine as a rational reconstruction of Thompson's matcher~\cite{thompson1968} in the light of Cox's elucidation as a virtual machine~\cite{coxregexptwo}. This machine uses a sequential schedule for incrementally evolving pointers, keeping a list of pointers that have been evolved already to prevent loops and search space explosion. However, our main interest is in performing this computation in parallel.

\section{Parallel lockstep semantics} \label{gpulockstep}
\begin{figure}[t]
\fbox{$M \longrightarrow M'$}
\[
\infern{ \proc_{1} \longrightarrow \proc_{2} }{
\proc_{1}  \parcomp \proc_{3} \longrightarrow \proc_{2} \parcomp \proc_{3}
}{\textsc{Par}}
\qquad
\infern{}{(( \ppre p\proc)  \parcomp \procsend p) \longrightarrow \proc}{\sendrule}
\]
\fbox{$M \astep a M'$}
\[
\infern{M' \not\equiv (\apre a M') \parcomp M''' \qquad M'\not\longrightarrow}
{(\apre a\proc_{1} \parcomp \ldots \parcomp \apre a\proc_{n} \parcomp
M'
)
\astep a (\proc_{1} \parcomp \ldots \parcomp \proc_{n})}{\sync}
\]
\caption{Process calculus}
\label{process}
\end{figure}

We now define an operational semantics where each pointer is given a dedicated thread for evolving it.
Our motivation is to leverage the large number of cores and hence threads available on GPUs. The semantics in this section is intended as an idealization of the implementation  described in Section~\ref{cudaimplementation} below, capturing the essentials of the computation while abstracting from implementation details.

To describe the parallel computation, we define a simple process calculus.
Its transition rules are given in Figure~\ref{process}. Most of our calculus is a subset of CCS~\cite{milnerccs}, with  one-to-one directional message passing and parallel composition. However, we also need an $n$-way synchronization with a synchronous transition inspired by Synchronous CCS~\cite{milnersynchrony}.

We let $\proc$ range over processes, $p$ over pointers that may be sent as asynchronous messages, and $a$ over input symbols, which may be used for $n$-way synchronisation. The syntax of processes is as follows:
\begin{eqnarray*}
M & \gramto& \procsend p
\\
&\mid& \proc \parcomp \proc
\\
&\mid& \ppre p\proc
\\
&\mid& \apre a\proc
\end{eqnarray*}

We impose some structural congruences $\equiv$, identifying terms up to associativity and commutativity of parallel compostion $\parcomp$.
Process transitions can be interleaved with rule \textsc{Par}.

We have CCS-style handshake communication in rule \sendrule. Here $\ppre p\proc$ receives the message $\procsend p$ and proceeds with $\proc$ afterwards. Note that receivers of the form $\ppre pM$ are not replicated (in the pi-calculus sense~\cite{milnerpibook}), so that each communication consumes the receiver. This behaviour is essential, as the processes we generate could become trapped in an infinite loop otherwise.


We also have an $n$-way synchronisation \sync. This rule is the most complex, and it is needed to implement matching to input once all pointers have been evolved. The idea is as follows:
\begin{itemize}
\item
The current process is factorized into those processes that are of the form $\apre a M_{j}$ and an $M'$ comprising everything else.
\item There are no further $\astep{}$ transitions inside $M'$, written as $M'\not\longrightarrow$.
\item If these conditions are met, then all the processes waiting to participate in an $n$-way  synchronization on $a$ are advanced in one synchronous step.
\item The remaining processes in $M'$ are discarded in the same step.
\end{itemize}

Rules in this style, in which a number of processes are advanced in a single step, are sometimes referred to as ``lockstep''~\cite{milnersynchrony}. Indeed, we use this rule to implement the lockstep matching of regular expressions in the sense of Thompson and Cox.
(In practice, this rule may require a little ad-hoc protocol to implement on a given architecture.)


We translate each expression pointer $p$ in the heap $\pi$ into a process $\sem p\,\pi$ as follows:
\begin{align*}
\sem{p}\,\pi &=
\ppre p(\procsend{q_{1}} \parcomp \procsend{q_{2}})
&&\mbox{if } \pi(p) = (q_{1} \mid q_{2})
\\[1ex]
\sem{p}\,\pi &=
\ppre p\procsend{q_{1}}
&&\mbox{if } \pi(p) = (q_{1} \bullet q_{2})
\\[1ex]
\sem{p}\,\pi &=
\ppre p(\procsend{q_{1}} \parcomp \procsend{q_{2}})
&&\mbox{if } \pi(p) = \kleene{q_{1}} \mbox{ and }\knode p =q_{2}
\\[1ex]
\sem p\,\pi &= \ppre p \procsend q
&&\mbox{if } \pi(p) = \epsexp \mbox{ and } \knode p = q
\\[1ex]
\sem p\,\pi &= \ppre p \apre a\procsend q
&&\mbox{if } \pi(p) = a \mbox{ and } \knode p = q
\end{align*}

Intuitively, for each internal node in the expression tree identified by the pointer $p$, we create a dedicated little process that listens on a channel uniquely corresponding to $p$. For simplicity, we use the same name for the channel as for the pointer. The process may be activated by messages $\procsend p$ sent to it, and it may send such messages itself. These messages trigger a chain reaction that evolve the current pointer set of a macro step. There is no need for these messages to be externally visible, as their only purpose is to wake up their unique recipient.
A process $\ppre pM$ listening for $\procsend p$ is consumed by the transition that receives the message.
Processes for nodes that point to input characters $a$ at the leaves of the expression tree use a different form of communication. All these nodes synchronize on the input symbol. The symbol $a$ is visible in the resulting synchronous transition step $\astep a$, because we need it to agree with the next input symbol.

If $\dom{\pi} = \{p_{1},\ldots, p_{n}\}$, we define the translation $\sem{\pi}$ as the translation of all its pointers:
\[
\sem{p_{1}}\,\pi \parcomp \ldots \parcomp \sem{p_{n}}\,\pi
\]


If the input string is not empty, let $a$ be the first character, so that $a\,w' = w$.
The parallel machine launches processes for all the nodes in the tree, and sends a message to the process for the root. The resulting process makes a number of asynchronous transitions, followed by a synchronous move for $a$:
\[
\sem\pi \parcomp \procsend p \longrightarrow \cdots \longrightarrow \;\astep a \proc
\]
All these steps together represent one macro step. The machine then repeats the above with the next symbol $a'$ and $\proc$
\[
\sem\pi \parcomp \proc  \longrightarrow \cdots \longrightarrow \;\astep {a'}\proc'
\]
The machine accepts if the remaining input is empty and the current process is of the form
\[
\procsend{\nullpointer} \parcomp \proc
\]

\begin{example}
\label{exampleprocess}
For $e = \kleene{\kleene{a}} b$, let $\pi$ and $p_{0}$ be such that $\pi,p_{0} \models e$. See Figure~\ref{figurepi} for the representation of $\pi$ as a tree with pointers. Translating the tree structure to parallel processes gives us:
\[
\sem{\pi} = (\ppre{p_{0}}\procsend{p_{1}}) \parcomp \ppre{p_{1}}(\procsend{p_{3}} \parcomp \procsend{p_{2}}) \parcomp \ppre{p_{2}}\apre{b}\procsend{\nullpointer} \parcomp \ppre{p_{3}}(\procsend{p_{4}} \parcomp \procsend{p_{1}}) \parcomp \ppre{p_{4}}\apre{a}\procsend{p_{3}}
\]
Assume an input string of $aab$. We have the pointer evolution as follows:
\begin{align*}
&\procsend{p_{0}} \parcomp \sem{\pi}\\
\longrightarrow &\procsend{p_{0}} \parcomp \ppre{p_{0}}\procsend{p_{1}} \parcomp \ppre{p_{1}}(\procsend{p_{3}} \parcomp \procsend{p_{2}}) \parcomp \ppre{p_{2}}\apre{b}\procsend{\nullpointer} \parcomp \ppre{p_{3}}(\procsend{p_{4}} \parcomp \procsend{p_{1}}) \parcomp \ppre{p_{4}}\apre{a}\procsend{p_{3}}\\
\longrightarrow &\procsend{p_{1}} \parcomp \ppre{p_{1}}(\procsend{p_{3}} \parcomp \procsend{p_{2}}) \parcomp \ppre{p_{2}}\apre{b}\procsend{\nullpointer} \parcomp \ppre{p_{3}}(\procsend{p_{4}} \parcomp \procsend{p_{1}}) \parcomp \ppre{p_{4}}\apre{a}\procsend{p_{3}}\\
\longrightarrow &\procsend{p_{3}} \parcomp \procsend{p_{2}} \parcomp \ppre{p_{2}}\apre{b}\procsend{\nullpointer} \parcomp \ppre{p_{3}}(\procsend{p_{4}} \parcomp \procsend{p_{1}}) \parcomp \ppre{p_{4}}\apre{a}\procsend{p_{3}}\\
\longrightarrow &\procsend{p_{3}} \parcomp \apre{b}\procsend{\nullpointer} \parcomp \ppre{p_{3}}(\procsend{p_{4}} \parcomp \procsend{p_{1}}) \parcomp \ppre{p_{4}}\apre{a}\procsend{p_{3}}\\
\longrightarrow &\apre{b}\procsend{\nullpointer} \parcomp \procsend{p_{4}} \parcomp \procsend{p_{1}} \parcomp \ppre{p_{4}}\apre{a}\procsend{p_{3}}\\
\longrightarrow &\apre{b}\procsend{\nullpointer} \parcomp \procsend{p_{1}} \parcomp \apre{a}\procsend{p_{3}}
\end{align*}
Since no more micro transitions are possible, we have reached the $n$-way synchronization point:
\[
\apre{b}\procsend{\nullpointer} \parcomp \procsend{p_{1}} \parcomp \apre{a}\procsend{p_{3}} \astep{a} \procsend{p_{3}}
\]
Now we feed the residual messages back into a fresh $\sem{\pi}$:
\begin{align*}
&\procsend{p_{3}} \parcomp \sem{\pi}\\
\longrightarrow &\procsend{p_{3}} \parcomp \ppre{p_{0}}\procsend{p_{1}} \parcomp \ppre{p_{1}}(\procsend{p_{3}} \parcomp \procsend{p_{2}}) \parcomp \ppre{p_{2}}\apre{b}\procsend{\nullpointer} \parcomp \ppre{p_{3}}(\procsend{p_{4}} \parcomp \procsend{p_{1}}) \parcomp \ppre{p_{4}}\apre{a}\procsend{p_{3}}\\
\longrightarrow &\ppre{p_{0}}\procsend{p_{1}} \parcomp \ppre{p_{1}}(\procsend{p_{3}} \parcomp \procsend{p_{2}}) \parcomp \ppre{p_{2}}\apre{b}\procsend{\nullpointer} \parcomp \procsend{p_{4}} \parcomp \procsend{p_{1}} \parcomp \ppre{p_{4}}\apre{a}\procsend{p_{3}}\\
\longrightarrow &\ppre{p_{0}}\procsend{p_{1}} \parcomp \ppre{p_{1}}(\procsend{p_{3}} \parcomp \procsend{p_{2}}) \parcomp \ppre{p_{2}}\apre{b}\procsend{\nullpointer} \parcomp \procsend{p_{1}} \parcomp \apre{a}\procsend{p_{3}}\\
\longrightarrow &\ppre{p_{0}}\procsend{p_{1}} \parcomp \procsend{p_{3}} \parcomp \procsend{p_{2}} \parcomp \ppre{p_{2}}\apre{b}\procsend{\nullpointer} \parcomp \apre{a}\procsend{p_{3}}\\
\longrightarrow &\ppre{p_{0}}\procsend{p_{1}} \parcomp \procsend{p_{3}} \parcomp \apre{b}\procsend{\nullpointer} \parcomp \apre{a}\procsend{p_{3}}\\
\astep{a} &\procsend{p_{3}}\\
\longrightarrow &\ldots\\
\astep{b} &\procsend{\nullpointer}
\end{align*}
Therefore, we have received a $\procsend{\nullpointer}$ while the input string has become empty, resulting in a successful match.
\end{example}

We need to prove that the construction above can correctly evolve and match any set of pointers.
Let $S=\{p_{1},\ldots,p_{n}\}\subseteq\dom\pi\cup\{\nullpointer\}$ be a set of pointers in the heap. We define
\[
\procsend S = \procsend{p_{1}} \parcomp \ldots \parcomp \procsend{p_{n}}
\]
to represent this set as a parallel composition of messages.

\begin{theorem}
\label{parallelmacrostep}
Let $S, S' \subseteq\dom\pi\cup\{\nullpointer\}$.
We have
\[
S \Astep{}\Astep a S'
\]
if and only if
\[
\procsend S \parcomp \sem\pi \astep{}^{*} \astep a \procsend{S'}
\]
Moreover, each $\astep{}$ transition sequence starting from $\procsend S \parcomp \sem\pi$ is finite.
\end{theorem}

Theorem~\ref{parallelmacrostep} assures us that the parellel operational semantics correctly implements the lockstep construction. The pointers $p$  in the tree, represented as processes $\procsend p$, are evolved in parallel. Although this evolution is non-deterministic, its end result is determinate. Moreover, the cycles in the pointer chain do not lead to cyclic processes looping forever, since each receiving process becomes inactive once the node has been visited.

The correctness proof of the parallel implementation relies on a factorisation of the processes into four components. At each step $i$, we have:
\begin{itemize}
\item A set $S_{i}$ of pointers, indicating nodes that should be evolved.
\item A heap of receivers $\pi_{i}\subseteq \pi$, representing nodes that have not been visited in the current macro step.
\item A set $E_{i}$ of evolved nodes, whose process representations are of the form ready to match a character.
\item A parallel composition $D_{i}$ of messages to nodes that have already been processed.
\end{itemize}

Let $E$ be a set of pointers $E=\{p_{1},\ldots,p_{n}\}$ such that $\pi(p_{j}) =a_{j}$ and
$\knode{p_{j}} =q_{j}$.
We write
\[
\waitchar E = \apre{a_{1}}q_{1} \parcomp\ldots\parcomp\apre{a_{n}}q_{n}
\]
We need to consider transition sequences of the form
\begin{eqnarray*}
&&\procsend{S_{0}} \parcomp \sem{\pi_{0}} \parcomp \waitchar{E_{0}} \parcomp D_{0}\\
& \astep{}&  \\
& \vdots&  \\
& \astep{}& \procsend{S_{n}} \parcomp \sem{\pi_{n}} \parcomp \waitchar{E_{n}} \parcomp D_{n}
\end{eqnarray*}
where $\pi_{0} =\pi$ and $E_{0}=\emptyset$. The invariant we need to establish for all transition steps consists of:
\begin{eqnarray*}
\evolve{S_{0}} &=& \evolve{(S_{i} \cap \dom{\pi_{i})}} \cup E_{i}
\\
\evolve{R_{i}} & \subseteq & \evolve{(S_{i} \cap \dom{\pi_{i}})} \cup E_{i}
\\
\{p \mid \exists D.D_{i} \equiv (\procsend p \parcomp D) \}
 & \subseteq & S_{i} \cup R_{i}
\end{eqnarray*}
where $R_{i} = \dom{\pi} \setminus \dom{\pi_{i}}$. The factorization of proceses at each step and the invariant are verified by case analysis on the kind of node $\pi(p)$  and hence the possible $\astep{}$ steps that its translation $\sem p\,\pi$ can make using the rules from Figure~\ref{process}.

In the final configuration we have $S_{n} \cap \dom{\pi_{n}} = \emptyset$. Hence,
\begin{eqnarray*}
\evolve{S_{0}} &=& \evolve{(S_{n} \cap \dom{\pi_{n})}} \cup E_{n}\\
&=& \evolve{\emptyset}  \cup E_{n}\\
&=& E_{n}
\end{eqnarray*}
Therefore, we have $\evolve{S_{0}} = E_{n}$, as required. From that configuration, there can only be an $\astep a$ transition, exactly matching the generic lockstep transition $S \Astep{}\Astep a S'$.

\section{Implementation on a GPU}
\label{cudaimplementation}

As a proof of concept, we have written a simple regular expression matcher where the evolution of pointers is performed in parallel on a GPU.\footnote{The code is available at \url{http://www.cs.bham.ac.uk/~hxt/research/regexp.shtml}.} Programming the GPU was done via CUDA~\cite{cudaintro}.
The main points are:
\begin{itemize}
\item
The regular expression is parsed, and the syntax tree nodes are packed into an array $d$. This array represents our heap $\pi$.
A second pass through the syntax tree performs the wiring of continuation pointers, corresponding to $\knode$.
\item
Two integer vectors $c$, $n$ of the same size as the regular expression vector above are created. Here a value of $t$ - the macro step count, on $c[i]$ implies that regular expression $d[i]$ is to be simulated within the current macro step. On the other hand a value of $-t$ on $c[i]$ implies that the corresponding regular expression has already been simulated for the current macro step. This protocol realizes the semantics of a process being consumed once it has received a message. The vector $n$ is used to collect those search attempts which are able to match the current input character. A value of $t + 1$ on $n[j]$ indicates that the regular expression $d[j]$ is to be simulated on the next macro step.
\item
Each regular expression node $d[i]$ is assigned a GPU thread. This GPU thread is responsible for conditionally simulating the regular expression $d[i]$ at each invocation (depending on $c[i]$ value). While simulating an expression, a GPU thread might schedule another GPU thread / expression $d[j]$ by setting $c[j]$ to $t$ (this could happen for an example in the case of $e = e_{1} \bullet e_{2}$).
Note that  one thread scheduling another thread via the $c$ vector corresponds to the sending of a message $\procsend p$ from one process to another.
\item
At each invocation of the GPU threads (called a \textit{kernel launch} in CUDA terminology), each thread which performs a successful simulation updates either of two shared flags which indicate if there were more threads activated on the $c$ or $n$ vectors during the current invocation.
A macro transition involves swapping the $c$ and $n$ vectors while incremeting the $t$ counter. It corresponds to the $n$-way synchronization transition.
\item
The initial state of the machine has only $d[0]$, the root node process, scheduled for simulation.
\end{itemize}
However, note that this description corresponds to a minimalistic GPU-based parallel lockstep machine and does not yet incorporate any optimizations from the literature~\cite{gpuirregular}, such as
\emph{persistent threads} and \emph{tasks queues}.

 \section{Conclusions}
\label{conclusions}

We have derived regular expression matchers as abstract machines. In doing so, we have used a number of concepts and techniques from programming language theory. The EKW machine zooms in on a current expression while maintaing a continuation for keeping track of what to do next. In that sense, the machine is a distant relative of machines for interpreting lambda terms, such as the SECD machine~\cite{landinmechanical} or the CEK machine~\cite{felleisensecd}.
On the other hand, regular expressions are a much simpler language to interpret than lambda calculus, so that continuations can be represented by a single pointer into the tree structure (or to machine code in Thompson's original implementation).
While the idea of continuations as code pointers is sometimes advanced as a helpful intuition, the representation of continuations in CPS compiling~\cite{appel} is more complex, involving an environment pointer as well.
To represent pointers and the structures they build up, we found it convenient to use a small fragment of separation logic~\cite{reynoldslicssep}, given by just the separating conjunction and the points-to-predicate. (They are written   as $\otimes$ and $\pi(p) =e$ above, to avoid clashes with other notation.)
A similar use of these connectives to describe trees in the setting of abstract machines was used in our earlier work on B+trees~\cite{btree}. Here we translate a tree-shaped data structure into a network of processes that communicate in a cascade of messages mirroring the pointers in the tree structure. The semantics of the processes is inspired by the process algebra literature~\cite{milnerccs,milnersynchrony,milnerpibook}. One reason why a process algebra is suitable for formalizing the lockstep construction with redundancy elimination is that receiving processes are eliminated once they have received a message; they are used linearly, and so are reminiscent of linearly-used continuations~\cite{LinUCHOSC}.

We intend to extend both the process algebra view and our CUDA implementation, while maintaining a close correspondence between them.
Regular expression matching is an instance of irregular parallel~\cite{gpuirregular} processing on a GPU, which presents some optimization problems.
At the moment, the parallel processing power of the GPU cores is not exercised, as each thread does little more than access the expression tree and activate threads for other nodes. We expect the load on the GPU cores to become more significant when more expensive constructs such as back-references (known to be NP-hard) are added to our matching language. It remains to be seen whether a GPU implementation will become more efficient than a sequential CPU-based one, particularly as the number of GPU cores continues to increase (it is currently in the hundreds of cores).
More generally, the operational semantics and abstract machine approach may be fruitful for reasoning about other forms of General Purpose Graphics Processing Unit (GPGPU) programming.


\end{document}